\documentclass{aa}
\input{epsf}
\usepackage[dvips]{graphicx}

\begin{document}

 \def \etal      {et al.\ }
\def \kev       {{\rm\ keV}}

% %   \thesaurus{     (11.03.1;  % Galaxies: clusters: general,
%                11.09.1;  % Galaxies: intergalactic medium,
%                12.03.3;  % Cosmology: observations,
%                12.04.1;  % Cosmology: dark matter,
%                12.12.1;  % Cosmology: large-scale structure oftheUniverse,
%                13.25.3)} % Xrays: general.

   \title{XMM-{\it Newton} Observation of the Coma Galaxy
   Cluster\thanks{Based on observations obtained with XMM-Newton, an 
      ESA science mission with instruments and contributions directly
funded by
    ESA Member States and the USA (NASA)}
\fnmsep
\thanks{EPIC was developed by the EPIC Consortium led by the Principal
Investigator, Dr. M. J. L. Turner. The consortium comprises the
following Institutes: University of Leicester, University of
Birmingham, (UK); CEA/Saclay, IAS Orsay, CESR Toulouse, (France);
IAAP Tuebingen, MPE Garching,(Germany); IFC Milan, ITESRE Bologna,
IAUP Palermo, Italy. EPIC is funded by: PPARC, CEA, CNES, DLR and ASI
}}

\subtitle{ The temperature structure in the central region }

\author{M. Arnaud \inst{1}, N. Aghanim\inst{2},R. Gastaud\inst{3},
   D.M. Neumann\inst{1}, D. Lumb \inst{4}, U. Briel \inst{5},
   B. Altieri\inst{6}, S. Ghizzardi\inst{7}, J. Mittaz\inst{8},
   T.P. Sasseen\inst{9},
    W.T. Vestrand\inst{10}}

\offprints{ marnaud@discovery.saclay.cea.fr} 

\institute{CEA/DSM/DAPNIA Saclay, Service d'Astrophysique, L'Orme des
Merisiers B\^at 709., 91191 Gif-sur-Yvette, France
\and IAS-CNRS,
Universit\'{e} Paris Sud, , B\^atiment 121, 91405 Orsay Cedex, France
\and CEA/DSM/DAPNIA Saclay, Service d'Electronique
    et d'Informatique, 91191 Gif-sur-Yvette, France
    \and Space Science Dept., European Space Agency, ESTEC Postbus 299,
2200AG Noordwijk, Netherlands
\and Max-Planck-Institut f\"{u}r
extraterrestrische Physik, D-85740 Garching, Germany
\and XMM-Newton
Science Operations Centre, ESA Space Science Department, P.O. Box
50727, 28080 Madrid, Spain
\and IFC/CNR, Via Bassini 15,
Milano,I-20133, Italy
\and Department of Space and Climate Physics,
UCL, Mullard Space Science Laboratory, Holmbury St.  Mary, Surrey, UK
\and University of California, Santa Barbara, CA 93110, USA
\and
NIS-2, MS D436,Los Alamos National Laboratory Los Alamos, NM 87545,
 USA }

%   \date{Received September 15, 1996; accepted March 16, 1997}
\date{Received 2 October  2000 / Accepted 2 November  2000}

\titlerunning{The temperature structure in the central region of
  Coma}
\authorrunning{Arnaud \etal}

%\begin{abstract}
\abstract{We present a temperature map and a temperature profile of
the central part ($r < 20'$ or 1/4 virial radius) of the Coma cluster. 
We combined 5 overlapping pointings made with XMM/EPIC/MOS and
extracted spectra in boxes of $3.5'\times3.5'$.  The temperature
distribution around the two central galaxies is remarkably homogeneous
($r<10'$), contrary to previous ASCA results, suggesting that the core
is actually in a relaxed state.  At larger distance from the cluster
center we do see evidence for recent matter accretion.  We confirm the
cool area in the direction of NGC 4921, probably due to gas stripped
from an infalling group.  We find indications of a hot front in the
South West, in the direction of NGC4839, probably due to an adiabatic
compression.   } \maketitle

%\end{abstract}

\def \msol      {{\rm\ M}_\odot}
\def \etal      {et al.\ }
\def \kev       {{\rm\ keV}}
\begin{table*}[t]
\caption[]{Observations}
\begin{tabular}{llllcccccc}
\hline
Obs. & Rev & RA. & DEC. &  MOS1 & MOS2&\multicolumn{4}{c}{MOS1\&2 counts}  \\
          & & (J2000.0)& (J2000.0)
&Exp.&Exp.&\multicolumn{2}{c}{[0.3-10] \kev}&\multicolumn{2}{c}{[5-10] \kev}\\
          & &  & & (ksec) & (ksec) &Source&Bkgd&Source&Bkgd\\
\hline
Pc
&86&$12^{h}59^{m}47^{s}$&$27\degr57\arcmin00\arcsec$&16.4&16.3&$7.70~10^5$&$4.2~
10^4$&$3.52~10^4$&$1.00~10^4$\\
P5
&86&$12^{h}59^{m}28^{s}$&$27\degr46\arcmin53\arcsec$&20.9&21.4&$6.63~10^5$&$5.5~
10^4$&$3.23~10^4$&$1.30~10^4$\\
P6
&93&$12^{h}58^{m}50^{s}$&$27\degr58\arcmin52\arcsec$&7.4&7.3&$1.81~10^5$&$1.9~10
^4$&$8.53~10^3$&$4.52~10^3$\\
P9
&93&$13^{h}00^{m}33^{s}$&$27\degr56\arcmin59\arcsec$&20.8&21.0&$7.03~10^5$&$5.4~
10^4$&$3.08~10^4$&$1.29~10^4$\\
P10
&98&$12^{h}59^{m}38^{s}$&$28\degr07\arcmin40\arcsec$&20.6&20.9&$5.26~10^5$&$5.4~
10^4$&$2.43~10^4$&$1.27~10^4$\\
\hline
\end{tabular}
\label{tab:obs}
\end{table*}

\section{Introduction}
Numerical simulations of cluster evolution (e.g. Evrard \cite{evrard};
Schindler \& M\"uller \cite{schindler}) suggest that the temperature
structure of the Intra-Cluster Medium (ICM) is a powerful indicator of
the evolutionary state of clusters.  In particular the accretion of a
sub-cluster, a common phenomenon in standard hierarchical formation
scenario, should manifest itself by characteristic features in the
temperature map, like heated gas between the two units just before the
collision.

Recent studies of the Coma cluster with the ASCA satellite (Donnelly
\etal \cite{donnelly}, Watanabe \etal \cite{watanabe}) revealed
complex temperature variations in this massive cluster.  They were
interpreted as indicative of recent mergers, confirming earlier
evidence based on optical dynamical studies (Colless \& Dunn
\cite{colless} and references therein) and X--ray morphological
analysis (Briel \etal \cite{briel1}; White \etal \cite{white},
Vikhlinin \etal \cite{vikhlinin1}, \cite{vikhlinin2}).  ASCA covered a
broad energy band, which is essential for precise temperature
estimate, but the observations suffered from a relatively large energy
dependent PSF. Therefore temperature structure determination with ASCA
might have been subject to systematic errors.  Furthermore the spatial
resolution was insufficient to resolve precisely the temperature
radial profile in the very core of clusters.

The EPIC instrument (Turner \etal \cite{turner}) on board XMM (Jansen
\etal \cite{jansen}) combines a high sensitivity with good spatial and
spectral resolution, on a wide energy range.  In this paper, we use
this unique capability to study the temperature structure in the
central ($\theta<20' = 0.78$ Mpc) region of Coma.  We present further
XMM results in two other papers of this issue: the large scale
morphology of Coma (Briel \etal \cite{briel2}) and the dynamics of the
infalling NGC 4839 group (Neumann \etal \cite{neumann3}).  In the
following, we assume $H_{0} = 50{\rm km/s/Mpc}$ and $q_{0}=0.5$.

\section{Data Analysis}

\subsection{The data}

The central part of Coma was observed with 5 overlapping pointings in
Full Frame mode with the EPIC/MOS camera (medium filter) and in
extended Full Frame mode with the pn camera.  As CTE correction in
this pn mode is still being studied, we considered only MOS data in
the present spectroscopic analysis.

We generated calibrated event files with SASv4.1, except for the gain
correction.  Correct PI channels are obtained by interpolating gain
values obtained from the nearest observations of the on board
calibration source.  Data were also checked to remove any remaining
bright pixels.  We excluded periods of high background induced by
solar flare protons.  We discarded all frames corresponding to a count
rate greater than 15 ct/100s in the $[10-12]$ \kev band, where the emission
is dominated by the particle induced background.  Finally, only events
in the nominal FOV are considered.

The central position of each pointed observation is listed in Table 1,
together with the revolution number and remaining observing time after
cleaning.

\subsection{Spatially resolved spectroscopy}

Spectra in various regions were considered to study temperature
variations.  Each pointing and each MOS camera are first treated
separately.

\subsubsection{Correction for vignetting effects}

The effective area at a given energy depends on position.  When
extracting the spectrum of a region, we weight each photon falling at
position $(x_{j},y_{j})$ of the detector and of energy $E_{j}$ by the
ratio of the effective area at that position, to the central effective
area (for this energy).  This weighting is taken into account in the
error estimate.  The `corrected' spectrum obtained is an estimate of
the spectrum one would get if the detector was flat.  A detailed
description of the method and of the vignetting calibration data used
are given in Arnaud et al.  (\cite{arnaud}).  Note that the vignetting
due to the RGA is included, but is assumed to be energy independent
(the variations are less than $1\%$ below 6 \kev).

\subsubsection{Background estimate}

We generated EPIC MOS background event files (one for each MOS camera)
by combining several high galactic latitude pointings.  The data are
cleaned for bright pixels, background flares and regions corresponding
to bright point sources are excluded.  The integrated exposure time is
94.3 ksec for MOS1 and 78.9 ksec for MOS2.  These event files can be
used for a proper estimate of the cosmic ray (CR) induced background
but not the X-ray background, which depends on pointing position and
filter.  However, bright cluster emission, like the one observed in
the Coma center, usually dominates the background except at high
energy (see also Arnaud \etal \cite{arnaud}).  We thus are mostly
sensitive to CR induced background.  Note also that the offset
pointings considered here always include the cluster peak emission, so
scattered light is not a problem.  The total estimated number of
source and background photons, in the $[0.3-10]\kev$ and
$[5-10]\kev$ energy ranges, are listed in Tab.\ref{tab:obs} for each
pointing.  Note the degradation of the S/N ratio at high energies as a
consequence of the very hard CR induced background.

It is known that the CR background changes slightly in the FOV. It is
thus better to consider the same extraction regions in detector
coordinates for the source and the background.  Furthermore, if one
wants to combine spectra of a given physical Coma region obtained from
different pointings, it is mandatory to define extraction regions in
sky coordinates.  To alleviate this practical problem, we simply
generated a specific background event file for each Coma pointing and
camera by modifying the sky coordinates of the background event file
using the aspect solution of the considered Coma observation.

For consistency, the background spectra were obtained using the same
correction method as used for the source.  The background component
induced by CR is not vignetted, but as we extract the background and
source spectra from the same region in detector coordinates the
correction factor is the same and does not introduces bias.

Further details on the characteristic of the EPIC/MOS background  and
subtraction method can be found in Arnaud \etal \cite{arnaud}.
\begin{figure}[t]
\epsfxsize=8.cm \epsfverbosetrue \epsfbox{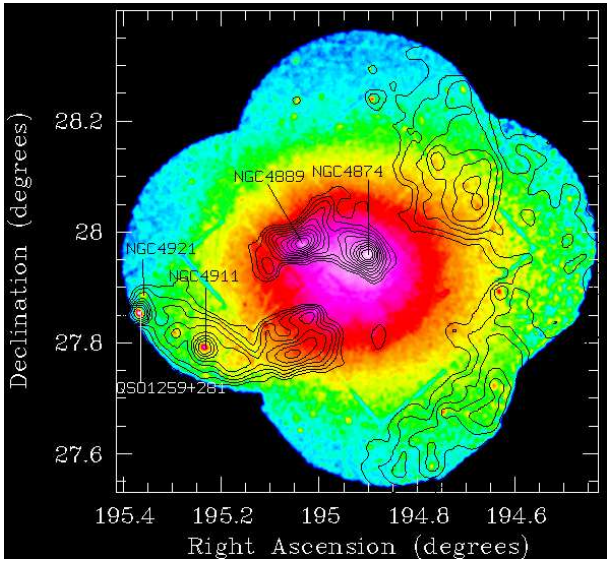} \caption[
]{{\footnotesize The EPIC/MOS mosaic image of the central region of
Coma (5 overlapping pointings) in the $[0.3-2]$  energy band.  The
iso contours are the residuals (in $\sigma$) after subtracting the best
fit 2--D $\beta$ model.  The step size is $4 \sigma$ and the lowest
iso-contour corresponds to $3 \sigma$ significance. The position of
the bright galaxies are marked. }}
\label{fig:fima}
 \end{figure}

\subsubsection{Spectra extraction and spectral fitting}

The source and background spectra of a given region (defined in sky
coordinates) are first extracted separately for each MOS camera and
each pointing, using the method described above.  As the spectra are
corrected for vignetting, the spectra of the same physical region
observed with different off-axis angle (from different pointings) and
different camera can be simply added to maximize the signal to noise
ratio.  The errors are propagated using quadratic summation.

Before model fitting, the source spectra are binned so that the S/N
ratio is greater than 3 $\sigma$ in each bin after background
subtraction.  The spectra are fitted with XSPEC using isothermal mekal
models (with fixed redshift $z=0.0231$).  Although the thermodynamic
state of the plasma could be more complex (e.g. see the isobaric
multiphase model of Nagai \etal \cite{nagai}), this is adequate to
study temperature spatial variations, the derived best fit temperature
being an estimate of the mean temperature in each projected region
considered.  Since the spectra are `corrected' for vignetting effects
we can use the on axis MOS response file, which is considered to be
the same for MOS1 and MOS2 (version v3.15).  Only data above 0.3 \kev
are considered due to remaining uncertainties in the MOS detector
response below this energy.  Unless otherwise stated errors are with a
$90\%$ confidence level.

\subsection{Imaging analysis}
The MOS mosaic image in the $[0.3-2]\kev$ energy band is presented in
Fig.~\ref{fig:fima}.  The count images for each camera and each
pointing are extracted using the weighting procedure described above
to correct for vignetting.  They are then projected on a common sky
reference axis, summed and divided by the mosaic exposure map (which
takes into account the exposure time and FOV coverage for each
pointing).  The images are not background subtracted.  In that energy
range and in this central area of the cluster, the particle background
is negligible for MOS.

\begin{figure}[t]
\epsfxsize=8.cm 
\epsfverbosetrue \epsfbox{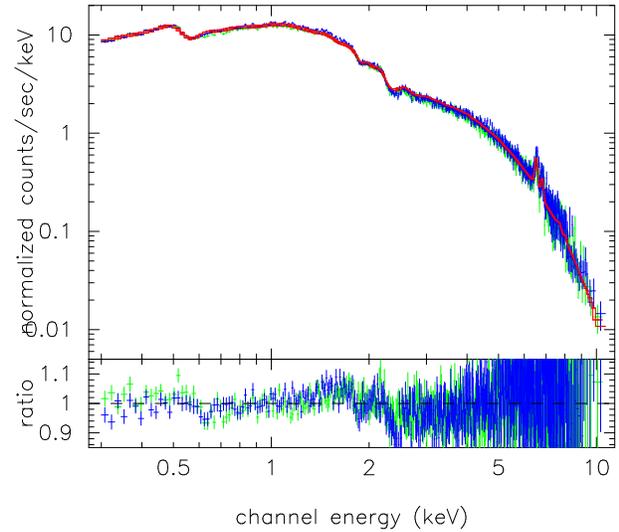} 
\caption[]{{\footnotesize EPIC/MOS1 (green) and EPIC/MOS2 (blue) spectra
extracted from within $10'$ in radius of the galaxy NGC 4874.
Red line: best fit redshifted isothermal model: ${\rm k}T= 8.25 \kev$
and
an abundance of $0.25$.  Bottom panel: residuals
between model and data.}}
\label{fig:fspecall}
 \end{figure}

\section{Results}

\subsection{Core morphology}

To identify significant substructures in the core, we fitted the MOS
image with a 2--D ellipsoidal $\beta$ model plus background and built
up the map of the residuals of the data over the best fit model.  The
method is discussed in detail in Neumann \& B\"ohringer
(\cite{neumann1}) and Neumann (\cite{neumann2}).  The iso-contours of
significance of the excess (number of $\sigma$ over background plus
cluster model) are overlaid on the MOS image in Fig.~\ref{fig:fima}.

We unambiguously confirm the excess emission around the two central
galaxies (NGC4874 and NGC 4889) and the tail of gas in the direction
of NGC4911, revealed by the wavelet analysis of Vikhlinin \etal
(\cite{vikhlinin2}).  However, contrary to Vikhlinin \etal result,
this filamentary structure does not seem to be directly connected to
the Coma center, but originates 0.5 Mpc South of it.  Note also that
part of the excess is due to resolved galaxy emission (NGC 4911, NGC
4921, QSO1259+281, and 5 other point sources).

Diffuse excess emission is clearly detected in the South-West in
particular in the direction of the NGC4839 group (see Briel \etal
\cite{briel2}, for full discussion).

\begin{figure}[t]
\epsfxsize=8.cm \epsfverbosetrue \epsfbox{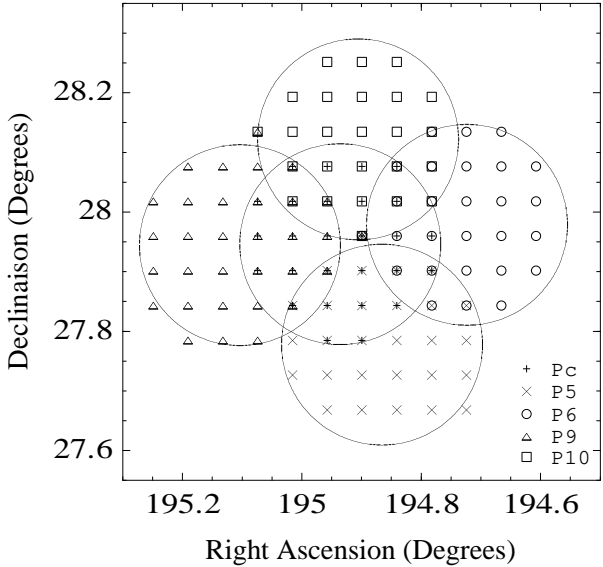} \caption[
]{{\footnotesize Position of the $3.5\times3.5'$ boxes, in which we
derive EPIC/MOS spectra.  A different symbol is used for each of the 5
pointed observations.  When the same region is observed in two or more
pointings, the corresponding spectra are summed (see text).  The
spectrum of each region is fitted with an isothermal model to build
the temperature map displayed in Fig.~\ref{fig:fkTmap}.  Circles:
central $\theta < 10'$ FOV of each pointed observation.} }
\label{fig:fbox}
\end{figure}

\subsection{Overall spectrum}

To compare with results obtained with other satellites we extracted
the overall MOS1 and MOS2 spectra from a circular region of 10 arcmin
in radius centered on NGC4874.  Only data from the central pointing
are used.  The data are fitted in the $[0.3-10] \kev$ range, with
independent normalizations for MOS1 and MOS2, and common temperatures
and abundances.  The spectra are plotted on Fig.~\ref{fig:fspecall},
the bottom panel gives the residuals.  Fixing the ($N_{\rm H}$ value
to the 21~cm value ($N_{\rm H}= 8.95\times 10^{19}~{\rm cm^{-2}}$, Dickey \&
Lockman \cite{dickey}), we obtain a best fit temperature of ${\rm k}T=
8.25\pm 0.10\kev$ and an abundance of $0.25 \pm 0.02$.  The reduced
$\chi^{2}$ is $1.38$ ($\chi^{2}=1457$ for $1058$ d.o.f).  The fit is
satisfactory.  The residuals are concentrated around the instrument
edges, with residual ratios between data and model of about $\pm 5\%$,
consistent with our present knowledge of the instrument response
(Fig.~\ref{fig:fspecall}).  If we let the $N_{\rm H}$ value free we get
${\rm k}T= 8.20\kev$, an abundance of $0.25$ and $N_{\rm H}=
9.4\pm0.9\times 10^{19}~{\rm cm^{-2}}$, in agreement with the 21cm value and
$\chi^{2}$ is unchanged.  In the following we thus fix the $N_{\rm H}$
value to the 21 cm value.

When fitted separately the MOS1 and MOS2 temperatures are consistent
and the relative normalization is 1.05.  In the following analysis we
will thus sum  MOS1 and MOS2 spectra.

The best fit overall temperature in this central ($R<10'$) region is
in remarkable agreement with the overall GINGA value of ${\rm k}T=
8.21\pm 0.09\kev$ (Hughes \etal \cite{hughes}) and is marginally
consistent with the ASCA value of $9.\pm 0.6 \kev$ ( Donnelly \etal
\cite{donnelly}) obtained for the central ($R<9'$) region.  Note that
the hard excess seen by SAX (Fusco-Femiano \etal \cite{fusco-femiano})
could not be seen by XMM, since it appears above 20 \kev.

\subsection{Temperature map}

\begin{figure}[t]
\center
\epsfxsize=7.cm \epsfverbosetrue \epsfbox{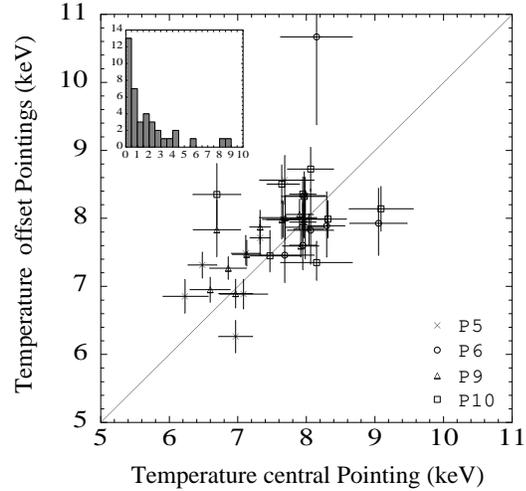} \caption[
]{{\footnotesize Temperature determined from the offset pointings
(P5, P6, P9 and/or P10) versus the temperature obtained, for the same
region, using the central pointing (Pc) data. Errors are at
$1\sigma$-level. The
insert shows the histogram of the differences in term of $\chi^{2}$
(see text for details). }}
\label{fig:fTonToff}
 \end{figure}

To study the cluster temperature structure, we next extracted spectra
in $3.5'\times3.5'$ contiguous regions in sky coordinates.  The box size
was chosen so that the two central galaxies fall approximately in the
center of a box, and that a sufficient S/N ratio is reached for each
box.  Circular regions ($20''$ in radius) around bright sources (in
particular NGC 4889 and NGC 4911) are excised from the boxes.  The
overall region considered for this spatially resolved spectroscopic
analysis is about $20'$ in radius. We only considered boxes at off-axis
angles smaller than $10'$ in each pointing (the
vignetting factor being more uncertain beyond this radius).  The
central $\theta<10'$ region of each pointing, delineated as circle,
and the central position of the various boxes (95 in total) are
plotted on Fig.~\ref{fig:fbox}.

\begin{figure}[t]
\epsfxsize=8.cm \epsfverbosetrue \epsfbox{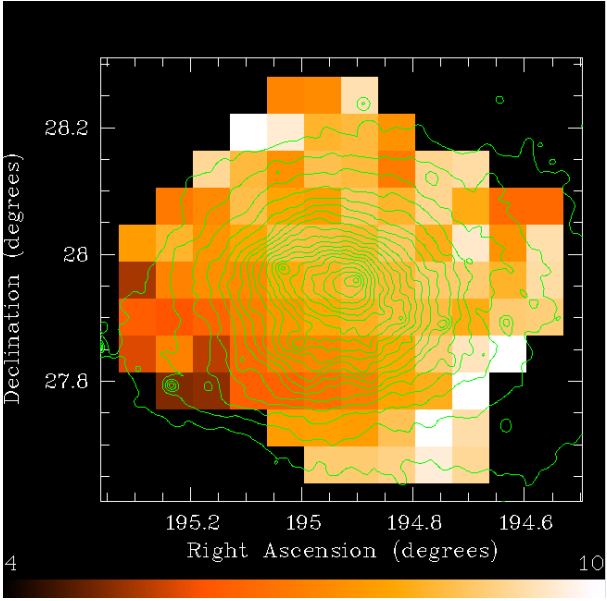} \caption[
]{{\footnotesize Color coded temperature map.  Note the hot front in
the south-west (white) and the cold region in the South-East
(brown/dark red).  The isocontours of the PN image in the $[0.3-2.]\kev$
band (Briel \etal \cite{briel2}) are superimposed.  The lowest contour
corresponds to $6.3\times 10^{-3}~{\rm ct/s/arcmin^{2}}$ and the step size is
$4.\times 10^{-3}~{\rm ct/s/arcmin^{2}}$}}
\label{fig:fkTmap}
 \end{figure}

The validity of our vignetting correction can be assessed by comparing
the fitted temperature of the same region in various pointings.  The
vignetting effect (decrease of effective area with off-axis angle)
increases with energy.  An understimate (overestimate) of this energy
dependence would yield to underestimate (overestimate) of the
temperature.  Since the various pointings of the same region
correspond to different off-axis angles, an improper vignetting
correction would translate in systematic differences between
temperature estimates for the same region.  Fig.~\ref{fig:fTonToff}
shows the temperatures determined from the offset pointings $T_{\rm
off}$ versus the temperature obtained, for the same region, using the
central pointing $T_{\rm c}$ (errors are at $1\sigma$-level).  The
insert shows the histogram of the differences in term of $\chi^{2} =
(T_{\rm off}-T_{\rm c})^{2}/(\sigma(T_{\rm off})^{2}+\sigma(T_{\rm
c})^{2})$ computed for each pair of measurements.  Three outliers
($\chi^{2} > 4$ or more than $2\sigma$ discrepancy in estimates) are
clearly apparent.  They correspond to the points at
(6.7,8.4),(6.5,7.3) and (7.6,8.5).  We do not see any particular
problem in the corresponding spectra (the statistic and fit are good)
and failed to find any obvious reason for the discrepancy.  However,
these outliers correspond to isolated regions (the agreement between
estimates is good in adjacent regions) and the overall $\chi^{2} = 44$
for 36 d.o.f is satisfactory when these outliers are excluded.  This
suggests that the vignetting correction is basically correct.  We thus
sum all spectra obtained for a given physical region and build up the
temperature map presented in Fig.~\ref{fig:fkTmap}.

\begin{figure}[t]
\epsfxsize=8.cm \epsfverbosetrue \epsfbox{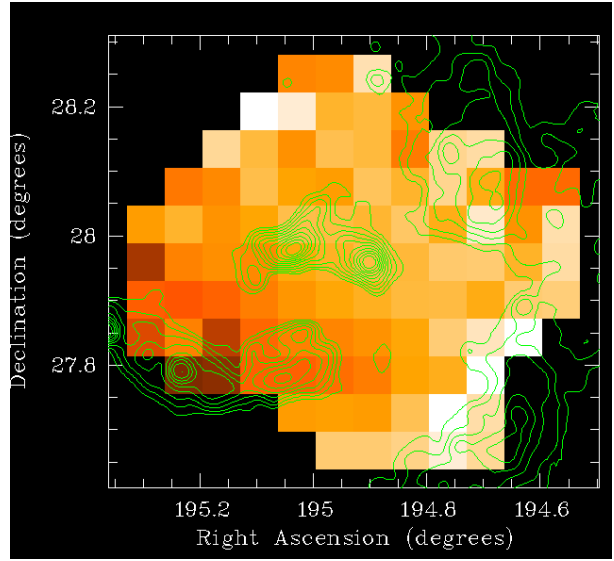} \caption[
]{{\footnotesize Excess emission over a $\beta$ model overlayed on the
temperature map.  Isocontours are as in Fig.~\ref{fig:fima}.  }}
\label{fig:fkTmapres}
 \end{figure}

\begin{figure}[t]
\epsfxsize=8.cm \epsfverbosetrue \epsfbox{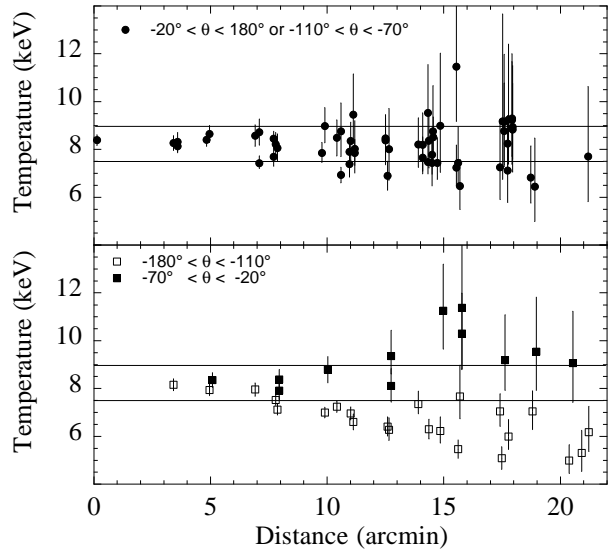} \caption[
]{{\footnotesize Temperature of each region of the temperature map
with $90\%$ confidence error bars.  The temperature is plotted versus
the distance to the brightest central galaxy NGC 4874 and the data are
splitted in three subsamples.  Bottom panel: the S-E sector
encompassing the filamentary structure towards NGC 4911 (open squares)
and the S-W sector along the direction towards NGC4839 group (filled
squares).  Top panel: the rest of the regions (filled circles).  The
angle (counterclockwise from west) defining the sectors are indicated
in the figure.}}
\label{fig:fkTbox}
 \end{figure}

There is no strong evidence of temperature variations, except for a
cold area in the South-East (contiguous regions colder than average)
and a hot area in the South-West.  It is instructive to compare these
temperature features with the X-ray image substructures
(Fig.~\ref{fig:fkTmapres}).  The S/E cold region in the temperature map
generally coincides with the filamentary substructure originating near
NGC 4911 and NGC 4921.  It includes the cold regions put into evidence
by Donnelly \etal (\cite{donnelly}) in that area (region 1 and part of
region 20 in their Fig.  2), that we thus confirm.  The hot regions in
the S/W appear as a hot front perpendicular to the direction
connecting the cluster center to the NGC4839 group, just ahead of the
excess emission in that direction.  This excess emission extends
somewhat further to the North, where no specific temperature feature
is apparent.  However, the temperature map is specially noisy in that
direction.  The statistical significance of the temperature variations
can be seen in Fig.~\ref{fig:fkTbox} where we plotted the temperature
of the various boxes versus their distance to NGC 4874.  We split the
data in three subsamples: i) one S-E sector encompassing the
filamentary structure towards NGC 4911 ii) one S-W sector along the
direction towards NGC4839 group iii) the rest of the sample.  The hot
front (k$T\sim 11 \kev$) in the S-W located at about $15'$ from the
center clearly stands out, as well as the colder region (k$T\sim 6
\kev$) beyond $10'$ in the S-E sector.  Otherwise the temperature does
not deviate significantly from the $8.25\pm0.75\kev$ temperature range
(less than $\pm 9\%$ variation).  In particular, we see no evidence of
the hot spot seen by Donnelly \etal (\cite{donnelly}) $3'$ north of the
NGC 4874 galaxy.  We further extracted the spectrum (using the central
pointing) corresponding to this hot ASCA region: from Fig 1 of
Donnelly \etal (\cite{donnelly}) we considered a rectangular region of
size $7.4\arcmin\times3.2\arcmin$ centered on
$\alpha=12^{h}59^{m}40^{s}, \delta=28\degr00\arcmin30\arcsec$.  We get
${\rm k}T= 8.4\pm 0.4\kev$ consistent with the mean value and
inconsistent with the ASCA value of Box 11 (${\rm k}T=
12.7^{+3.6}_{-2.0} \kev$).

\begin{figure}[t]
\epsfxsize=8.cm \epsfverbosetrue \epsfbox{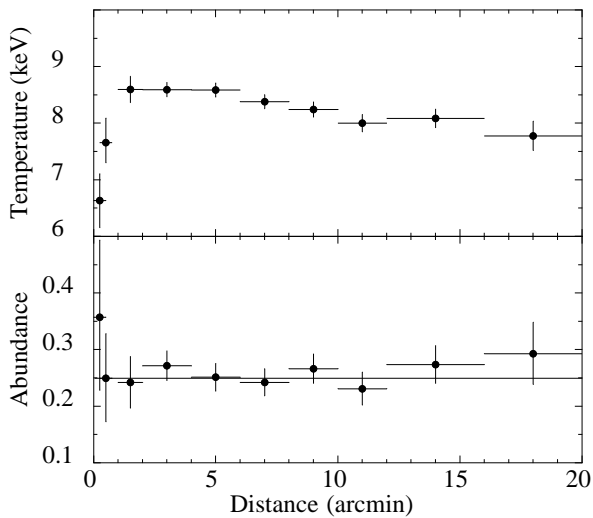}
\caption[ ]{{\footnotesize Radial temperature (top panel) and
abundance (bottom panel) profiles. The rings are centered on NGC
4874. }}
\label{fig:fkTprof}
 \end{figure}

\subsection{Temperature profile}
We further extracted spectra in concentric rings centered on NGC 4874
(all the data are summed).  The region around NGC 4889 ($40''$ in radius)
and bright point sources were excluded.  The temperature and abundance
profiles are shown on Fig.~\ref{fig:fkTprof}.

\section{ Discussion and Conclusion}

The dynamical state of the core of Coma has been much debated, in
particular the nature of the merging unit(s) and their link with the
dominant cluster galaxies.  If there is a consensus that a merging
group is associated with NGC4889, the situation of NGC4874 is less
clear (see in particular Colless \& Dunn \cite{colless}; Donnelly
\etal \cite{donnelly}).  As already proposed by Colless \& Dunn
(\cite{colless}), our data suggest that NGC4874 is simply the central
galaxy of the main Coma cluster, rather than being associated with a
second subgroup in an early merging stage with the main cluster (as
proposed by Donnelly \etal \cite{donnelly}).  Several facts support
this picture.  First the remarkably homogeneous temperature
distribution within the central $\theta<10'$ region suggests the gas
in that region is basically in a relaxed state.  Second, there is no
obvious evidence of a third peak in between NGC4889 and NGC4874, which
would be associated with the `true' cluster center.  Actually, apart
from the excess around NGC4889, the X--ray morphology can be
classified as an offset-center cluster morphology (variation of
isophote centroid with scale).  The excess (size $\sim 3'$ in radius)
around NGC4874, when subtracting a $\beta$ model (representative of
the large scale morphology) is a natural consequence for this type of
morphology.  Moreover, part of the excess is certainly due to the
contribution of the halo of the galaxy itself.  The very significant
drop in temperature within $1'= 40 {\rm kpc}$ of NGC4874
(Fig.~\ref{fig:fkTprof}) is natural in that context, as well as the
increase in abundance, which could be due to an enriched ISM.

It might be surprising that substructures survive in the gas density
distribution (excess around NGC4889 and centroid shift for the main
cluster) while the temperature distribution appears homogeneous.  We
must first emphasize that the spatial resolution and accuracy of the
temperature and imaging data is not comparable.  However, our results
may also indicate that the gas simply follows the dark matter
distribution.  Dark matter substructures can survive for a long time
after mergers, as indicated by high resolution simulations (Moore
\etal \cite{moore}).  High resolution hydrodynamic simulation, and
more sophisticated morphology analysis, are essential to better
understand this issue.

At larger scale we do see evidence of recent merger activity.  The
cold filamentary structure in the South-East can be naturally
explained by a merging group (see Vikhlinin \etal \cite{vikhlinin2}
and Donnelly \etal \cite{donnelly}).  The position and extent of the
cold substructure and the core properties outlined above suggest that
the merging group is associated with NGC4911 and NGC4921, rather than
being due to gas stripped from a group centered on NGC4874 as proposed
by Donnelly \etal (\cite{donnelly}).  Note that an excess in the
galaxy distribution is also observed around NGC4921/NGC4911 (Mellier
\etal \cite{mellier}).

Our analysis revealed for the first time a hot front in the
South-West, just ahead of the excess emission that we see at the edge
of the MOS mosaic and which extends further away towards NGC4839 (see
Briel \etal \cite{briel2}).  It is situated roughly at the boundary of
the group associated with this galaxy, as defined from the optical
(Colless \& Dunn \cite{colless}, Fig.  9) and is perpendicular to the
direction connecting the center of Coma and NGC4839.  This temperature
structure is likely to be due to adiabatic compression, caused by the
infall of matter associated with the NGC4839 group.  The feature we
find is indeed very similar to the one displayed in Fig.6c of
Schindler \& M\"uller (\cite{schindler}), where at this time of the
merger event no accretion shock has yet formed.  To definitively
characterise the feature we need to know the temperature structure at
larger radii, to quantify the transition conditions.  It has already
been noted that the NGC4839 group is located along the large scale
filament connecting Coma and A1367 (e.g West \etal \cite{west}).  It
is commonly thought that clusters form preferentially through
anisotropic accretion of sub-clusters along large scale filaments. 
Our finding supports this scenario.  The merger activity in that
direction, particularly interesting for our understanding of cluster
formation, is further discussed in Briel \etal (\cite{briel2}) and
Neumann \etal (\cite{neumann3}).

Except for the very center as discussed above, the abundance is
constant and the temperature radial profile is very weakly decreasing
with radius.  The slight drop beyond $10'$ is likely to be due to the
cold S/W structure.  The implications of this profile for the
distribution of dark matter in the core will be studied in a
forthcoming paper.

\begin{acknowledgements}
We would like to thank J. Ballet for support concerning the SAS
software and J.-L. Sauvageot for providing the gain correction.  We
thank S.Schindler and the anonymous referee for useful comments, which
improved the paper.
\end{acknowledgements}

\end{document}